\documentclass{ptephy_v1}

\preprintnumber{XXXX-XXXX}

\newcommand{\degree}{${}^\circ\mathrm{}$}
\newcommand{\degreeC}{${}^\circ\mathrm{C}$}

\title{Development of New Tracking Detector with Fine-grained Nuclear Emulsion for sub-MeV Neutron Measurement}

\author{T. Shiraishi\thanks{Corresponding author}}
\affil{Department of Physics, Toho University, Chiba 274-8510, Japan \email{takuya.shiraishi@sci.toho-u.ac.jp}}

\author{I. Todoroki}
\affil{Graduate School of Science, Nagoya University, Aichi 464-8602, Japan}
\author[1,3]{T. Naka}
\affil{Kobayashi-Maskawa Institute, Nagoya University, Aichi 464-8602, Japan}
\author[2]{A. Umemoto}
\author[2]{R. Kobayashi}
\author{O. Sato}
\affil{Institute of Materials and Systems for Sustainability, Nagoya University, Aichi 464-8602, Japan}

\usepackage{bm}
\usepackage{graphicx}
\usepackage[left]{lineno}
\usepackage{comment}
\usepackage{subfig}

\graphicspath{{./figure/}}

\begin{document}

\begin{abstract}
In this study, we have developed a new sub-MeV neutron detector that has a high position resolution, energy resolution, directional sensitivity, and low background. The detector is based on a super-fine-grained nuclear emulsion, called the Nano Imaging Tracker (NIT), and it is capable of detecting neutron induced proton recoils as tracks through topological analysis with sub-micrometric accuracy. We used a type of NIT with AgBr:I crystals of (98 $\pm$ 10) nm size dispersed in the gelatin. First, we calibrated the performance of NIT device for detecting monochromatic neutrons with sub-MeV energy generated by nuclear fusion reactions, and the detection efficiency for recoil proton tracks of more than 2 $\mu$m range was consistently 100\% (the 1 $\sigma$ lower limit was 83\%) in accordance with expectations by manual based analysis. In addition, recoil energy and angle distribution obtained good agreement with kinematical expectation. The primary neutron energy was reconstructed by using them, and it was evaluated as 42\% with FWHM at 540 keV. Furthermore, we demonstrated newly developed an automatic track recognition system dedicated to the track range of more than a few micrometers. It achieved a recognition efficiency of (74 $\pm$ 4)\%, and recoil energy and angle distribution obtained good agreement with manual analysis. Finally, it indicated the very high rejection power for $\gamma$-rays.
\end{abstract}

%\subjectindex{sub-MeV Neutron; Fine-grained Nuclear Emulsion; Dark Matter}
\subjectindex{}

\maketitle

%\linenumbers

\section{Introduction}
Neutron measurement is an important subject in a wide range of studies such as low background experiments in the particle physics (e.g., dark matter search \cite{DAMA,XENON}, neutrinoless double beta-decay search \cite{KamLAND,CANDLES}), neutron imaging from nuclear fusion or fission reactors \cite{Fusion1,Fusion2,Fission}, and neutron radiography \cite{Radiography}. In these measurements, the $^3$He proportional counter and the liquid scintillator are often utilized. For example, the $^3$He proportional counter is well suited for the measurement of thermal neutrons, however, it has no energy resolution (surely, no direction sensitivity) because it detects protons by the neutron absorption reaction $^3$He(n,p) after deceleration by the moderator. For the liquid scintillator, $\gamma$-rays become the background in a large $\gamma$/neutron ratio environment; in particular, it difficult to discriminate between neutrons and $\gamma$-rays in the sub-MeV or less region. Each of these methods has some disadvantages, and furthermore, they are not suitable for neutron imaging or directional detection because of the lack of spatial resolution. In this study, we propose a super-fine-grained nuclear emulsion, called the Nano Imaging Tracker (NIT) \cite{NIT1,NIT2}, as a new neutron detector with measurement capability of sub-MeV or more. This detector is expected well working in, for example, very high $\gamma$/neutron environment, neutron imaging and so on with both energy and spatial resolution.

In nuclear-emulsion-based neutron measurement, hydrogen, which is one of its components, is the main target for neutron detection. In an early approach to neutron detection with a nuclear emulsion by Y. Nomura et al. \cite{Iguchi}, they manually measured neutrons in the several-MeV region by tracking recoil protons $>$10 $\mu$m with an optical microscope. The automatic analysis was developed by S. Machii et al. \cite{Machii}, and the measurement of surface environmental neutrons was performed at energies greater than 2 MeV.

For neutrons in the sub-MeV region, the recoil proton tracks are several micrometers, and therefore sufficient studies have not been conducted up to now. In the directional dark matter search experiment \cite{NEWSdm}, which is the primary objective of NIT detector development, detection technology for 100 nm tracks has been implemented \cite{DFT}, and an adequate environment for track detection and analysis $>$1 $\mu$m has been provided.
The underground environment for the dark matter search has the neutron flux of approximately 5 orders of magnitude less than the $\gamma$-ray flux, therefore the properties of neutrons in such environment have not been well understood yet, and high $\gamma$-ray separation capability is required for rare event search experiments working in such environment. The NIT detector is expected to have an extremely high $\gamma$-ray rejection power compared to existing neutron detectors owing to its detection principle, not only high precision tracking of protons induced by neutrons.

In this study, our primary contribution is a new technique for measuring sub-MeV neutrons and evaluating its performance. In section \ref{sec:setup}, we describe the details of the NIT device used in this study. To verify the detection capability for sub-MeV neutrons, an experiment with monochromatic neutrons from the fusion reaction was conducted, and the performance of its principle of sub-MeV neutron detection is evaluated in section \ref{sec:performance}. In section \ref{sec:auto}, we propose an automatic analysis method with an optical microscope system for more practical use, and evaluate its implementation and performance. Further, we evaluate the rejection power for $\gamma$-rays as a background event, and examine its feasibility in the future high $\gamma$/neutron ratio environment.

\section{Experimental Setup}
\label{sec:setup}

\subsection{Detector}
\label{subsec:detector}

\begin{figure}[h!]
  \centering
  \includegraphics[width=12cm,bb=0 0 800 250]{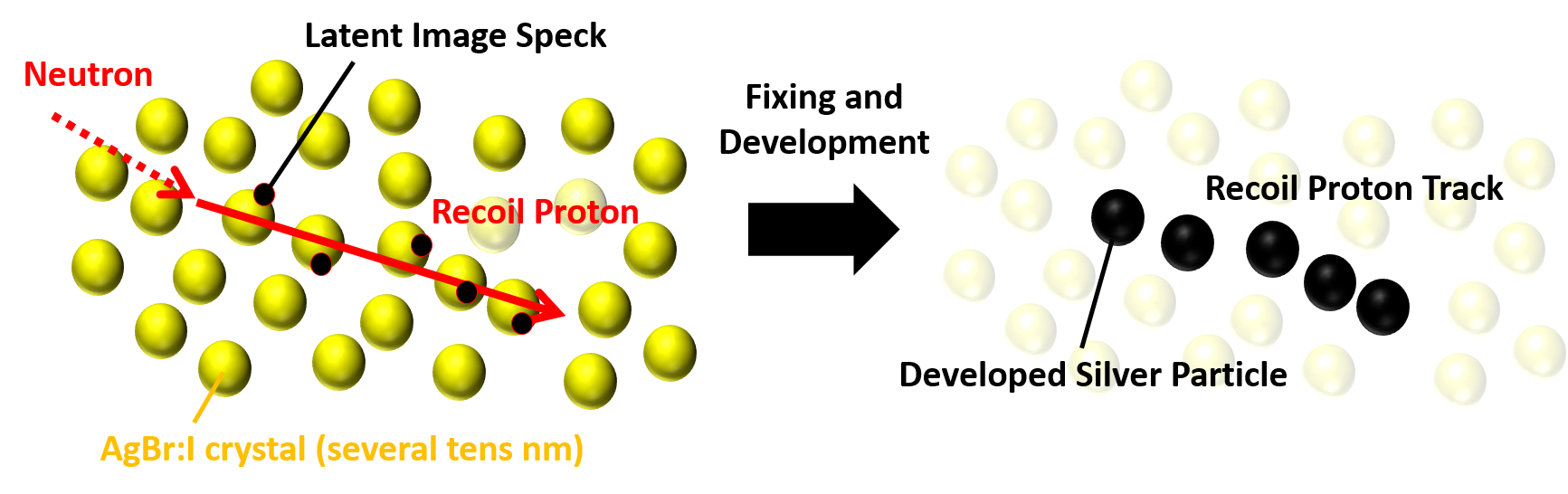}
  \caption{Schematic diagram of neutron detection principle by the nuclear emulsion. A track is formed as a series of developed silvers along a neutron induced proton recoil in the NIT.}
  \label{fig:principle}
\end{figure}

The NIT is a charged particle tracking detector in which silver iodobromide (AgBr:I) crystals of several tens of nanometers are dispersed with high density in a gelatin and polyvinyl alcohol medium with a stable crystal size distribution of approximately 15\%. Each AgBr:I nanocrystal behaves as a semiconductor sensor for charged particles. When electrons are excited by the charged particles, they repeatedly trap surface defects and desorb by thermal energy. A silver atom is created by attracting interstitial silver ions while electrons remain in the trap. When this reaction occurs more than four times at the same trap site, silver clusters called latent image specks (LISs) are created. LISs grow to approximately the original crystal size scale by chemical development, as shown in Fig. \ref{fig:principle}. These grown LISs, which form a track, are called developed silver particles.

In general, because a larger crystal exhibits a higher reflection intensity of developed silver particles with an epi-illumination microscope, it is easier to recognize a track. In this study, a NIT with a AgBr:I crystal size of (98 $\pm$ 10) nm, mass density of (3.2 $\pm$ 0.2) g/cm$^{3}$, and crystal density of approximately 1000 crystals/$\mu$m$^{3}$ was produced to consider proton tracking accuracy and $\gamma$-ray rejection power.
The fractional mass of hydrogen contained in the NIT is (1.75 $\pm$ 0.30)\%. Using a slide glass (size 76 $\times$ 26 mm$^{2}$, thickness 1 mm) as a base material, a NIT of approximately 35 $\mu$m thickness was applied to prepare a sample.

In fact, this initial condition of the NIT leads to low sensitivity to high dE/dx particles such as recoil protons. This makes track reconstruction difficult because the developed silver grain density is too low. Therefore, we always apply chemical sensitization treatment to such an NIT device for more efficient formation of latent image specks. The chemical sensitization treatment adopted was Halogen-Acceptor sensitization \cite{HA}. Here, sodium sulfite of 3.97 mol/L was utilized, and the dried NIT film after pouring was soaked in that solution for 15 min at 20.0 \degreeC, and dried again.

Development treatment after exposure was performed with MAA-1 Developer \cite{MAA}, which is a very popular developer in photographic science for surface development of the AgBr:I crystal, and it is the current standard development used for the NIT device. The treatment condition was set as development for 10 min at 5.0 \degreeC. The details are shown in the Appendix. The undeveloped AgBr:I crystals are dissolved by the fixing treatment. As a result, the NIT thickness up on analysis is shrunk (0.53 $\pm$ 0.02) times the original thickness. This value is called the shrinkage factor.

\subsection{Monochromatic sub-MeV Neutron Exposure}
\label{subsec:exposure}

Fast neutrons are primarily generated by nuclear fusion reactions or the spontaneous fission of radioisotope sources such as $^{252}$Cf. In particular, fusion reactions can generate neutrons with monochromatic energy, and therefore they are often used to calibrate detectors. In this study, the following two types of fusion reactions (endothermic reactions, whose targets are lithium and tritium) provided by the National Institute of Advanced Industrial Science and Technology (AIST) as the standard neutron field \cite{AIST} were used to verify the detection of monochromatic neutrons with sub-MeV energy:
\begin{eqnarray}
{}^{7}Li + p + 2.3MeV \longrightarrow  {}^{7}Be + n + 565keV, \\
{}^{3}T + p + 1.7MeV \longrightarrow  {}^{4}He + n + 880keV.
\end{eqnarray}

\begin{figure}[h!]
  \includegraphics[width=10cm,bb=0 0 600 400]{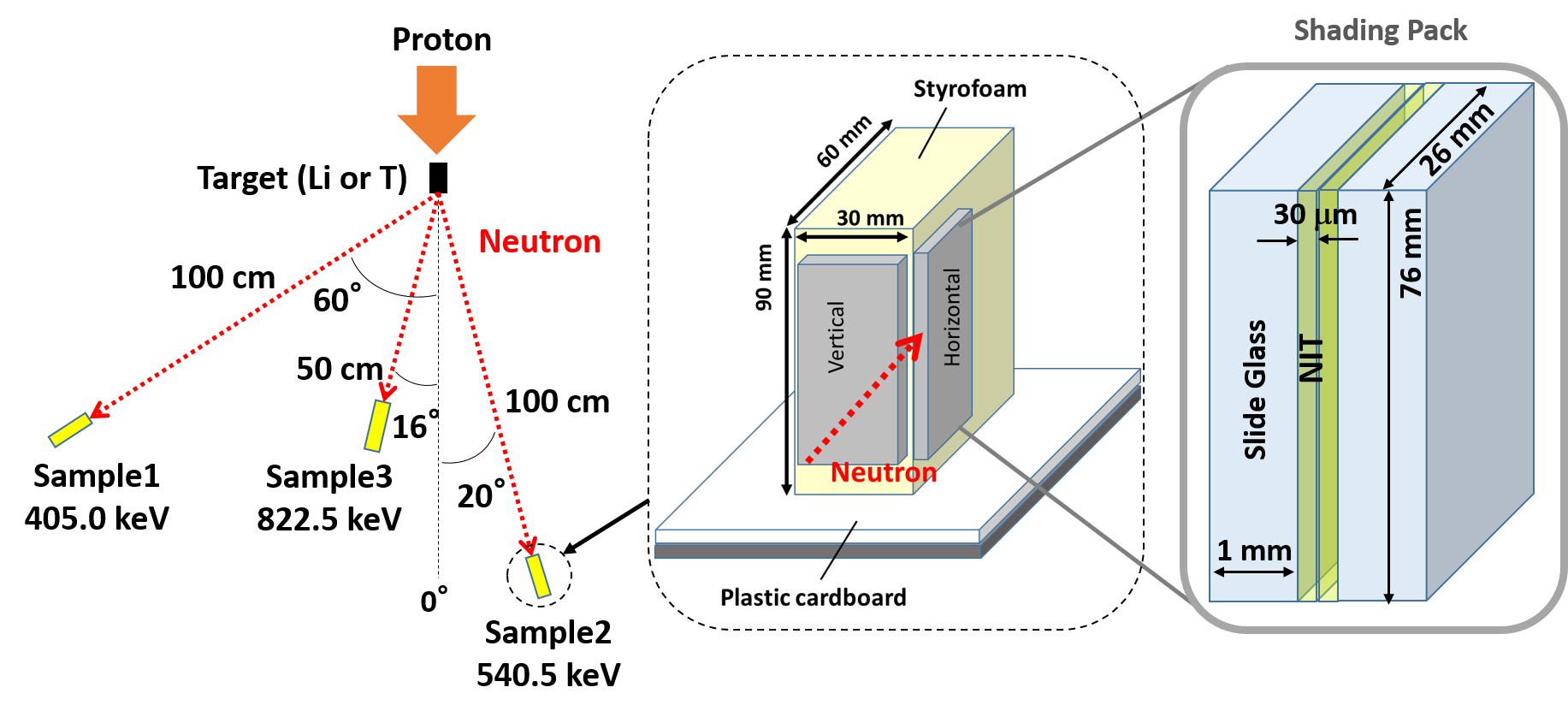}
  \caption{Experimental setup for monochromatic sub-MeV neutron exposure at AIST.}
  \label{fig:setup}
\end{figure}

The emitted neutron energies are respectively 565 and 880 keV at an angle of 0\degree. The energy can be adjusted by changing their emission angle with respect to the sample. The setup of neutron exposure at AIST is shown in Fig. \ref{fig:setup}. Table \ref{tab:sample} shows the neutron flux and the emitted energy($E_{n,AIST}$) for each sample. Here, the value of flux is the evaluated ones from calibrated neutron monitoring system and beam current by the AIST, and the $E_{n,AIST}$ are the calculated ones from beam angle and the proton beam energy.  In addition, an unexposed sample was prepared as the reference for comparison with the exposed samples.

\begin{table}[htb]
\centering
\caption{Sample conditions of monochromatic sub-MeV neutron exposure at AIST. Neutron energy and flux were calculated and measured by AIST, respectively \cite{AIST}.}
\begin{tabular}{|c||c|c|c|c|c|c|} \hline
& Nuetron & Distance & Angle & ${E}_{n, AIST}$ & Flux & Exposure time \\
& source & (cm) &  & (keV) & (n cm$^{-2}$ s$^{-1}$) & (hour) \\ \hline
Sample1 & Li(p,n)Be & 100 & 60\degree & 405.0 $\pm$ 9.3 & 361 $\pm$ 12 & 5.68 \\
Sample2 & Li(p,n)Be & 100 & 20\degree & 540.5 $\pm$ 12.7 & 791 $\pm$ 26 & 5.68 \\
Sample3 & T(p,n)He & 50 & 16\degree & 822.5 $\pm$ 16.7 & 908 $\pm$ 30 & 6.92 \\ \hline
Reference & - & - & - & - & 0 & 0 \\
sample &  &  &  &  &  & \\ \hline
\end{tabular}
\label{tab:sample}
\end{table}

\subsection{Optical Microscope Readout System}
\label{subsec:optics}

For the analysis of recoil proton tracks in the NIT, we used the Post Track Selector (PTS) \cite{PTS-2}, which is an automatic scanning system with an epi-illumination optical microscope. As for the development status of the PTS system, PTS-2 has been reported up to now \cite{PTS-2}, and a faster system is currently being developed.
In PTS data acquisition, as shown in Table \ref{tab:specification}, the field of view (FOV) of 112 $\mu$m $\times$ 60 $\mu$m is taken 116 times by a CMOS camera every 0.3 $\mu$m thickness, which approximately corresponds to the depth of field, and high-resolution tomographic images (three-dimensional image) of 112 $\mu$m $\times$ 60 $\mu$m $\times$ 35 $\mu$m are obtained. These image data are transferred from the camera to the SDRAM on the PC at high speed, after which high-speed scanning can be performed by image processing on the SDRAM (online analysis). When performing manual analysis or offline analysis, tomographic images are saved for each FOV.

\begin{table}[htb]
\centering
\caption{Specification of current Post Track Selector (PTS) system.}
\begin{tabular}{|c|c|} \hline
  Objective lens & N.A. 1.45, 100$\times$ \\
  Light source & LED, 455 $\pm$ 27 nm, 17 W \\ \hline
  Pixel resolution & 0.055 $\mu$m \\
  Number of pixels & 2048 $\times$ 1088 \\
  Camera frame rate & 300 fps \\
  FOV & 112 $\mu$m $\times$ 60 $\mu$m \\ \hline
  Layer pitch & 0.30 $\mu$m \\
  Number of layers per FOV & 116 \\ \hline
%  Data-taking speed & 0.73 s/view \\
%                    & 43 g/year \\ \hline
\end{tabular}
\label{tab:specification}
\end{table}

\section{Performance Evaluation by Manual Analysis}
\label{sec:performance}

The recoil proton tracks in the NIT can be easily recognized with the human eye from the tomographic images acquired by PTS (manual analysis). First, to demonstrate the principle of sub-MeV neutron detection, we evaluated the performance of manual analysis as benchmark data, and compared it with the simulation to evaluate its reliability.

\subsection{Neutron Energy Reconstruction}
\label{subsec:e_rec}

\begin{figure}[h!]
  \begin{minipage}[b]{0.4\linewidth}
    \centering
    \subfloat[]{\includegraphics[width=9cm,bb=50 0 850 400]{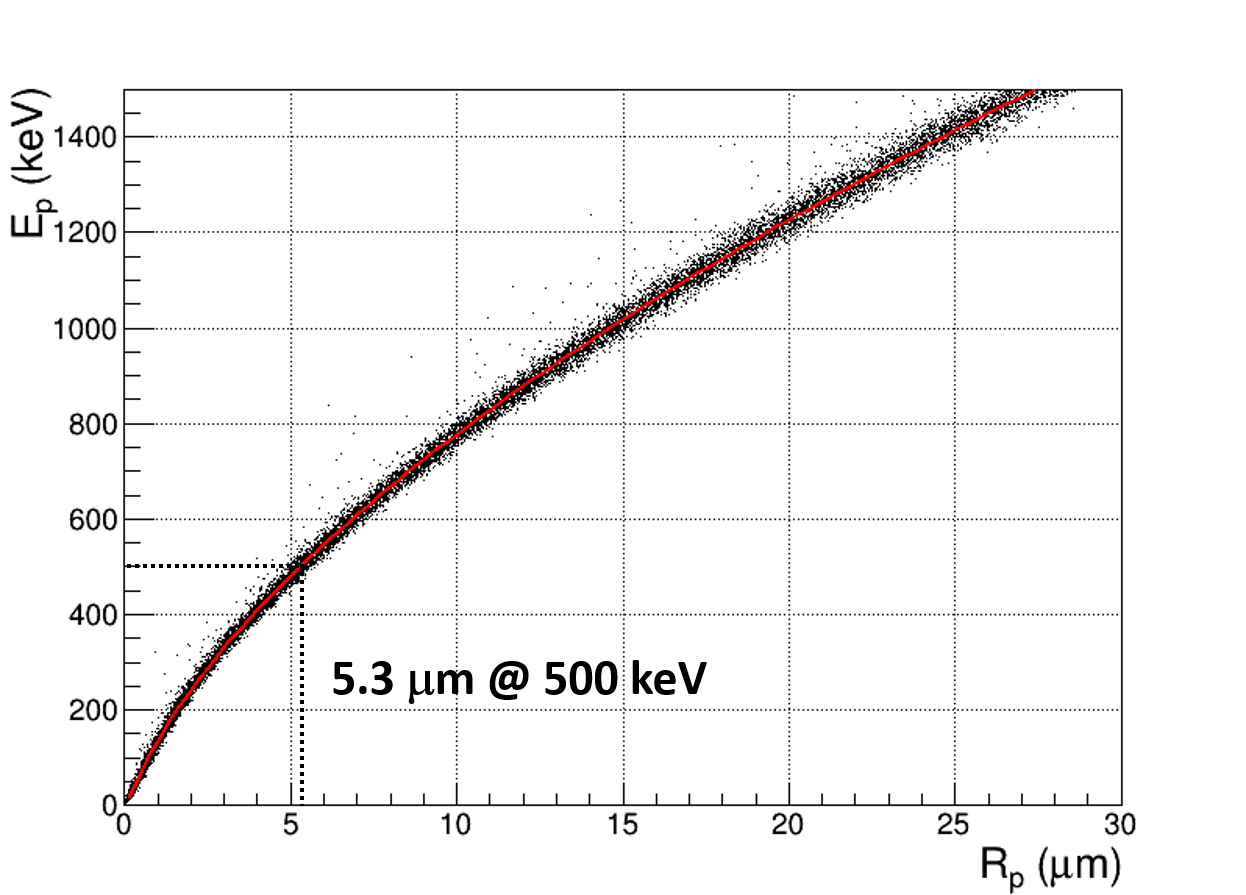}\label{fig:rp_ep}}
  \end{minipage}
  \begin{minipage}[b]{0.6\linewidth}
    \centering
    \subfloat[]{\includegraphics[width=9cm,bb=20 0 620 250]{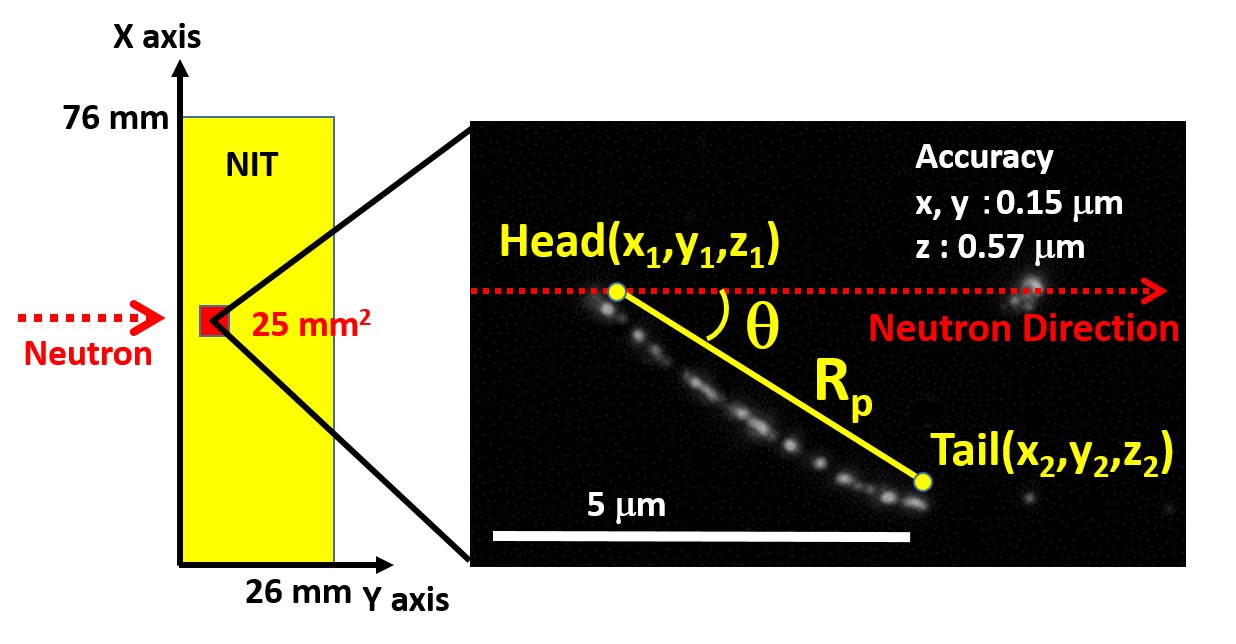}\label{fig:p_recoil}}
  \end{minipage}
  \caption{(a) Correlation of proton energy (Ep) and range (Rp) from the simulation in the NIT with Geant4; red line is fitted by Eq. \ref{eq:proton_energy}. (b) An example of proton recoil event found by manual analysis in the NIT.}
  \label{fig:proton}
\end{figure}

The information obtained in this measurement is the range of the recoil proton track (${R}_{p}$ [$\mu$m]) and the scattering angle from the neutron beam axis (${\theta}_{p}$). The relation between ${R}_{p}$ and the kinetic energy of the proton (${E}_{p}$ [keV]) is shown in Fig. \ref{fig:rp_ep} from the simulation in the NIT with Geant4. In this measurement, Eq. \ref{eq:proton_energy} was used as an approximate curve when obtaining the proton energy:
\begin{equation}
  \label{eq:proton_energy}
  {E}_{p} = 41.6 + 527 \times {R}_{p}^{1/2} - 432 \times {R}_{p}^{1/3}.
\end{equation}

If the arrival direction of the neutrons is known, the neutron energy (${E}_{n, mes}$) can be reconstructed from classical kinematics as Eq. \ref{eq:neutron_energy}:
\begin{equation}
  \label{eq:neutron_energy}
  %{E}_{n} = \frac{4Mm}{{(M+m)}^{2}} {E}_{p} {cos}^{2}{\theta}_{p},
  {E}_{n, mes} = \frac{{E}_{p}} {{cos}^{2}{\theta}_{p}}.
\end{equation}
Because the NIT can also accurately determine the scattering angle ${\theta}_{p}$, the neutron energy ${E}_{n, mes}$ can be reconstructed for each event. Reconstructing the energy of monochromatic neutrons from the signals obtained by manual analysis provides objective evidence of sub-MeV neutron detection.

\begin{figure}[h!]
  \begin{minipage}[b]{0.5\linewidth}
    \centering
    \subfloat[]{\includegraphics[width=10cm,bb=0 0 800 450]{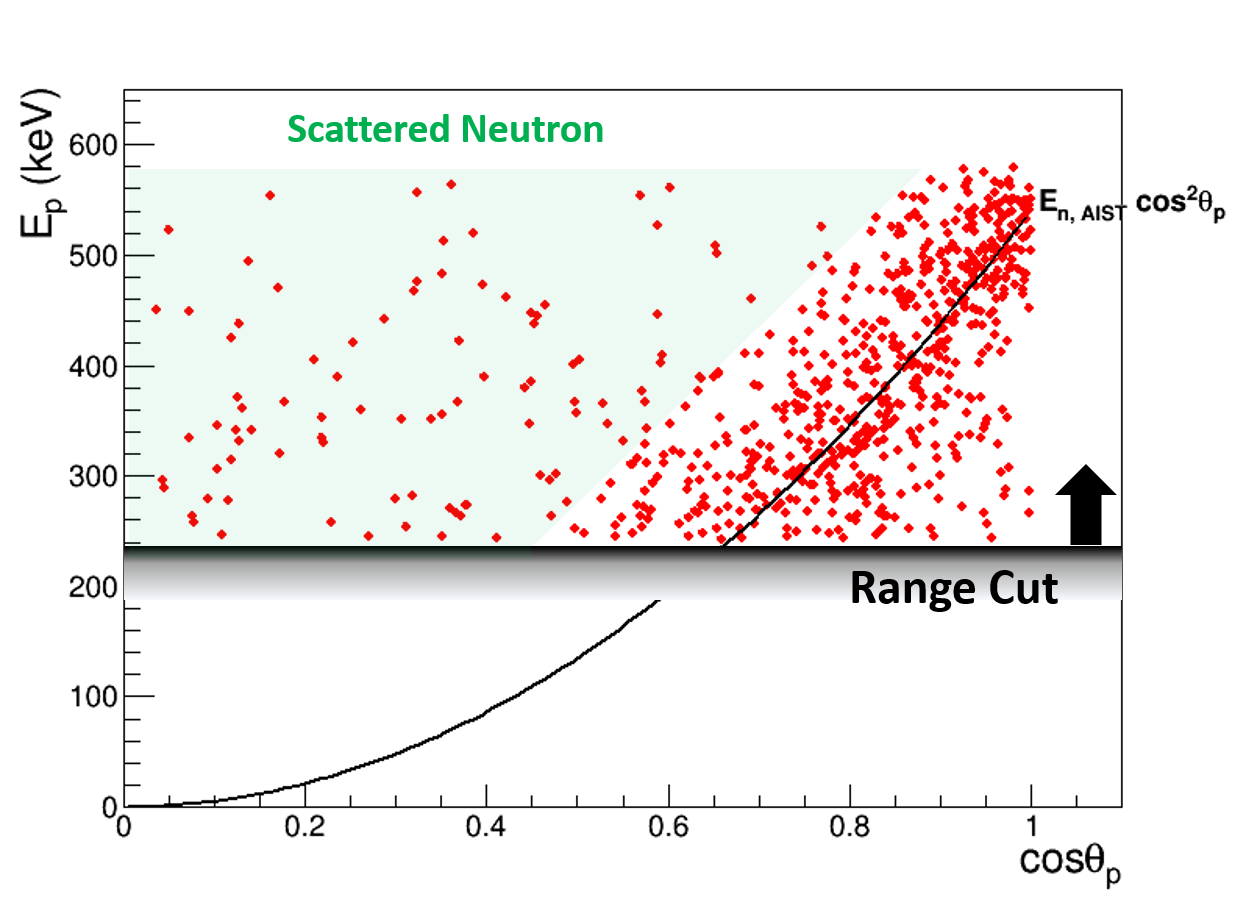}\label{fig:e_cos_sample2}}
  \end{minipage}
  \begin{minipage}[b]{0.5\linewidth}
    \centering
    \subfloat[]{\includegraphics[width=10cm,bb=0 0 800 450]{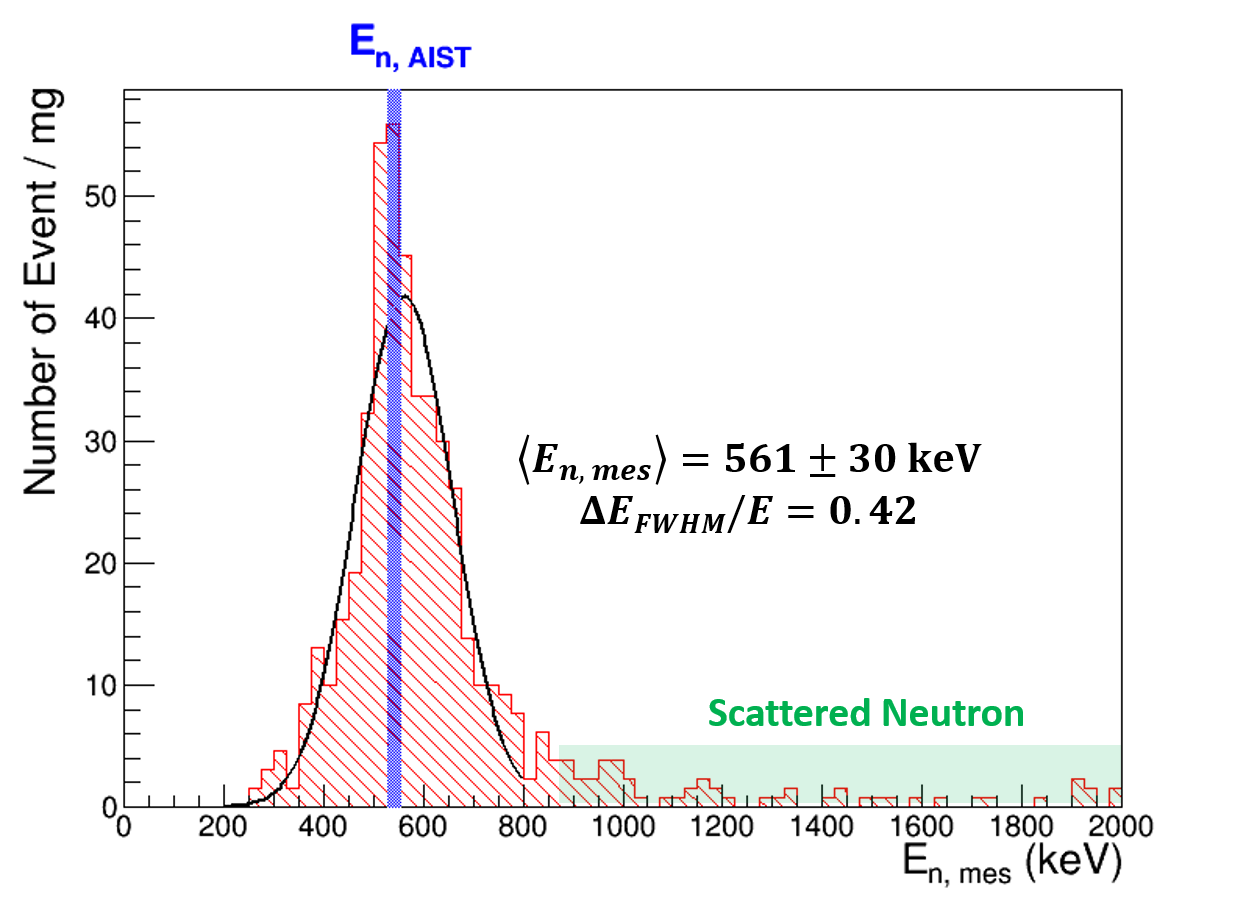}\label{fig:e_rec_sample2}}
  \end{minipage}
  \caption{Results of manual analysis of 1.3 mg volume for Sample2. (a) Recoil energy vs. scattering angle. (b) Distribution of reconstructed neutron energy by using proton recoil energy and angle.}
  \label{fig:result_sample2}
\end{figure}

The proton tracks were searched by manual analysis from the tomographic image captured by the PTS, the three-dimensional coordinates of the head and tail points were measured as shown in Fig. \ref{fig:p_recoil}, and the range ${R}_{p}$ and scattering angle ${\theta}_{p}$ were calculated from the line segment. The position accuracy of the developed silvers that form the track is approximately 0.15 $\mu$m for X and Y, and 0.57 $\mu$m for Z.
%Because the range of tracks that go out of the analysis volume cannot be calculated accurately, only the tracks whose start point and end point are within the analysis area were considered (Fiducial Volume Cut, F.V.C.).\textcolor{blue}{Volumeカットを具体的に書く。} 
In the emulsion layer of approximately 30 $\mu$m, 3 $\mu$m each from the upper and lower surfaces were cut, and approximately 24 $\mu$m inside was used for analysis as a fiducial volume. This cut is applied to prevent recoil proton tracks from going out of the analysis area because the range cannot be calculated accurately, and the external $\alpha$-ray background.
Fig. \ref{fig:result_sample2} shows the results of 586 events of recoil proton tracks obtained by manual analysis of a 1.3 mg volume for Sample2. This analysis applied a proton track range cut of more than 2 $\mu$m, which is equivalent to a proton energy of 240 keV. We did not analyze below the 2 $\mu$m range because of the lack of device purification.

\begin{figure}[h!]
  \centering
  \includegraphics[width=10cm,bb=0 0 750 500]{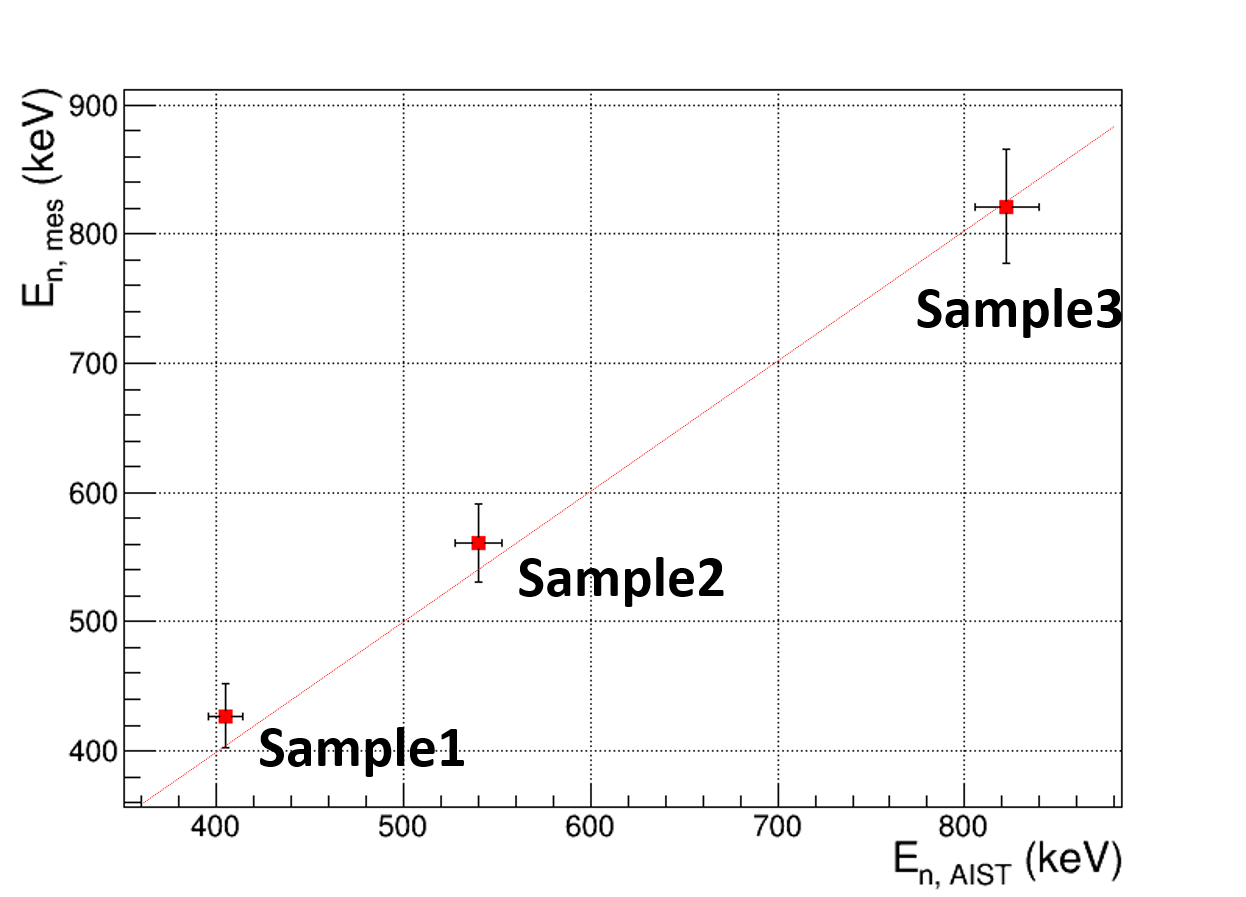}
  \caption{Comparison of ${E}_{n, AIST}$ and reconstructed neutron energies for all samples.}
  \label{fig:e_rec_all}
\end{figure}

Fig. \ref{fig:e_cos_sample2} shows the scatter plot of the scattering angle and recoil energy of the detected protons. Most of them are distributed around the curve corresponding to Eq. \ref{eq:neutron_energy}, and the components outside this curve are expected to be protons recoiled by neutrons that arrive at the NIT after scattering by the surrounding materials. The distribution of ${E}_{n, mes}$ reconstructed by Eq. \ref{eq:neutron_energy} is shown in Fig. \ref{fig:e_rec_sample2}.
The mean value of the measured ${E}_{n, mes}$ in Sample2 is 561 $\pm$ 30 keV, which includes errors of statistics, the shrinkage factor, and the NIT density. It is almost the same as the neutron energy ${E}_{n, AIST}$ exposed to Sample2, and the energy resolution of FWHM is 42\%. In addition, as is evident from the results of performing the same analysis on Sample1 and Sample3 exposed with different monochromatic energies, as shown in Fig. \ref{fig:e_rec_all}, both of them can be reconstructed accurately.

\subsection{Comparison with Monte Carlo Simulation}
\label{subsec:comparison}

\begin{figure}[h!]
  \centering
  \includegraphics[width=7cm,bb=0 0 500 300]{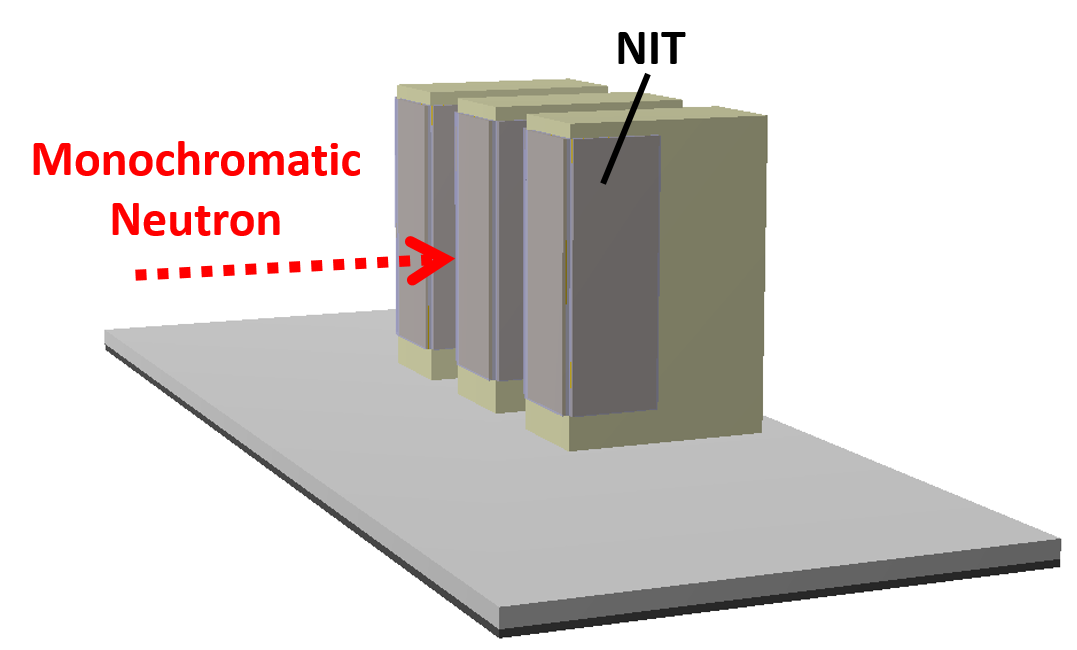}
  \caption{Geometry of MC simulation with Geant4 for the monochromatic neutron measurement at AIST.}
  \label{fig:geometry}
\end{figure}

To evaluate the reliability of the manual analysis results, a comparison with Monte Carlo (MC) simulation was made using Geant4. A geometry very similar to that shown in Fig. \ref{fig:setup} was created for the MC simulation, as shown in Fig. \ref{fig:geometry}.

The number of detected recoil proton tracks is 451 $\pm$ 87 event/mg (the statistical error is 4.1\%, systematic errors are included as the NIT density error 6.5\%, the neutron dose error is 3.3\%, the analysis volume error is 3.8\% due to the shrinkage factor, and the hydrogen content error is 17.1\%), and the expected number of recoil proton tracks by the MC simulation is 442 $\pm$ 16 event/mg. The detection efficiency of the manual analysis is 100\% consistent (the 1 $\sigma$ lower limit was 83\%).

\begin{figure}[h!]
  \begin{minipage}[b]{0.5\linewidth}
    \centering
    \subfloat[]{\includegraphics[width=10cm,bb=0 0 800 400]{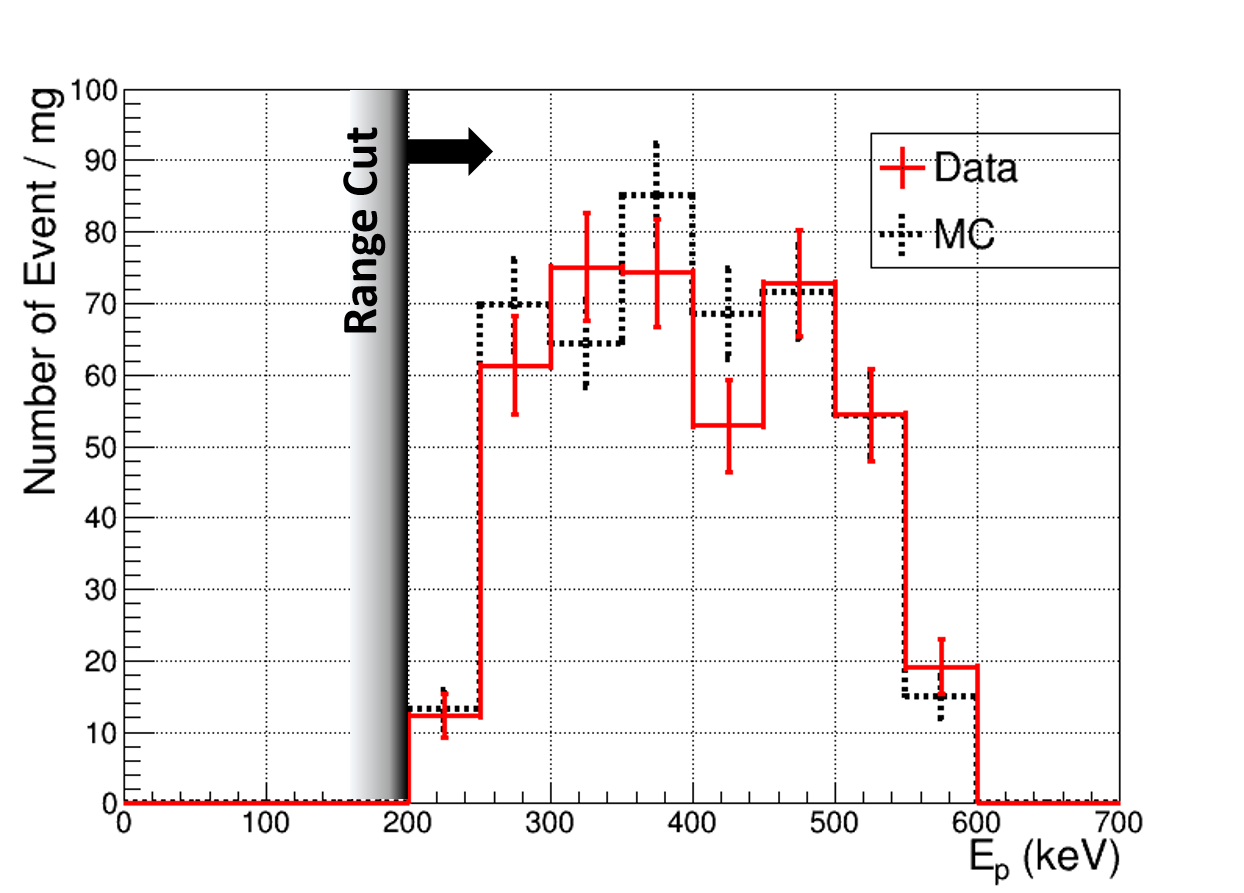}\label{fig:recoil_e}}
  \end{minipage}
  \begin{minipage}[b]{0.5\linewidth}
    \centering
    \subfloat[]{\includegraphics[width=10cm,bb=0 0 800 400]{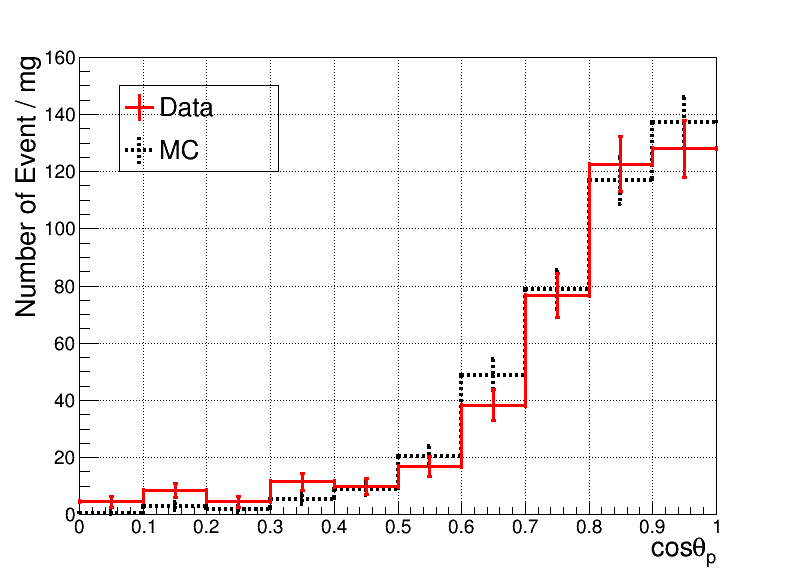}\label{fig:cos}}
  \end{minipage}
  \caption{Comparison of the selected event data with MC simulation. (a) Recoil energy (Sample2), and (b) Scattering angle (Sample2).}
  \label{fig:comparison}
\end{figure}

Fig. \ref{fig:comparison} compares the proton recoil energy and scattering angle for data and MC. ${E}_{p}$ and ${\theta}_{p}$ in the MC simulation are calculated by smearing the head and tail points of the track with the measurement errors. Both the calculated values are in good agreement with the kinematical prediction. This result shows that the reliability of manual analysis is high, there is no detection bias, and almost no noise is present.

\section{Automatic Analysis}
\label{sec:auto}

As verification of the principle of detection in section \ref{sec:performance}, sub-MeV neutron detection by manual analysis of the NIT is shown to have high detection efficiency and energy resolution for proton tracks $>$2 $\mu$m. In manual analysis, the recoil proton tracks are searched from the tomographic images, and the track range and angle are measured manually, which limits the analysis speed, and this analysis can be applied up to the 10 mg scale. In this section, we developed an automatic analysis system to improve the event selection capability by automatically recognizing tracks and measuring the track range and angle according to the image-data-taking speed of the PTS. This system aims to trigger event candidates and automatically measures the track range and angle, and assumes that manual analysis is performed after automatic analysis as the final confirmation.

\subsection{Development of Chain Analysis}
\label{subsec:readout}

\begin{figure}[h!]
  \centering
  \includegraphics[width=12cm,bb=0 0 800 400]{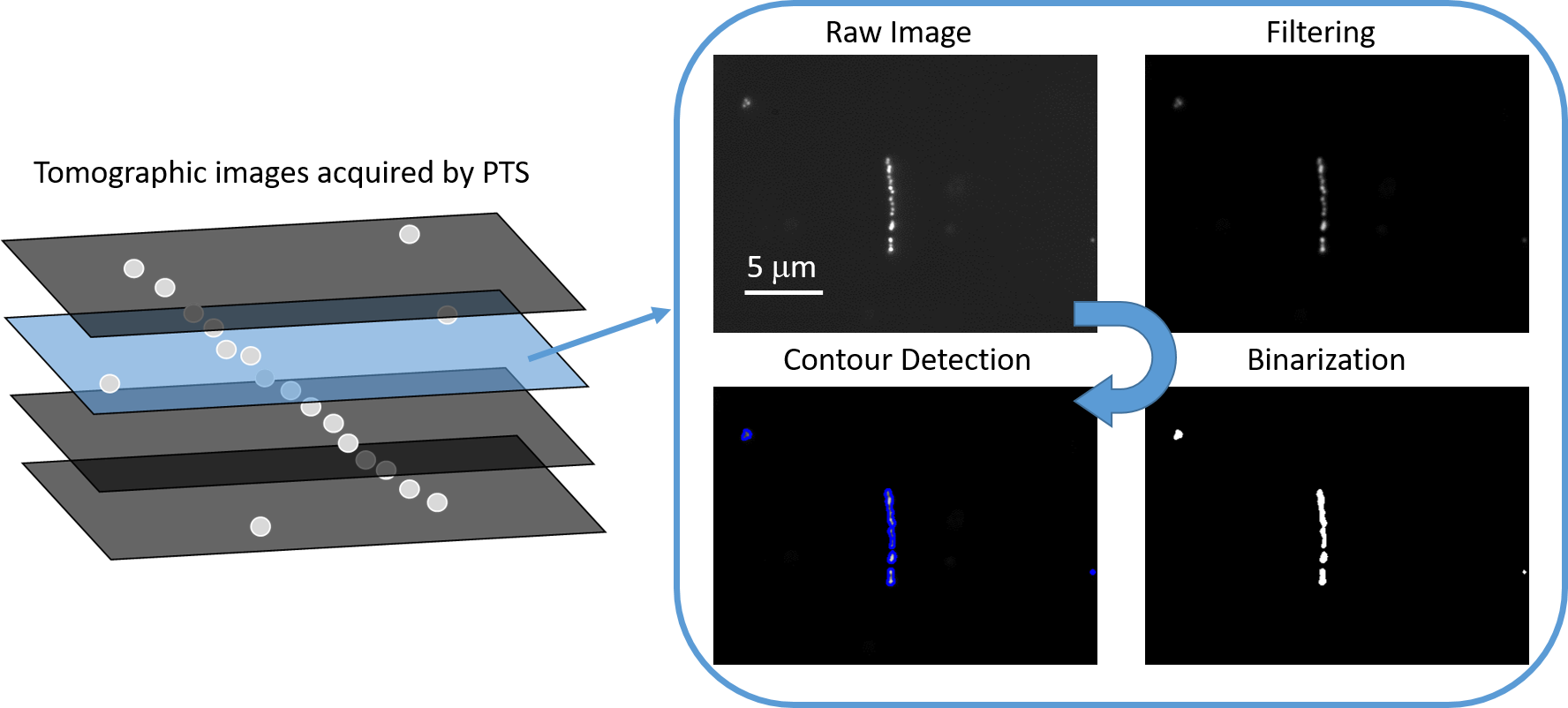}
  \caption{Schematic image and example data for filtering and track contour extraction of chain analysis.}
  \label{fig:filtering}
\end{figure}

\begin{figure}[h!]
  \centering
  \includegraphics[width=12cm,bb=0 0 900 600]{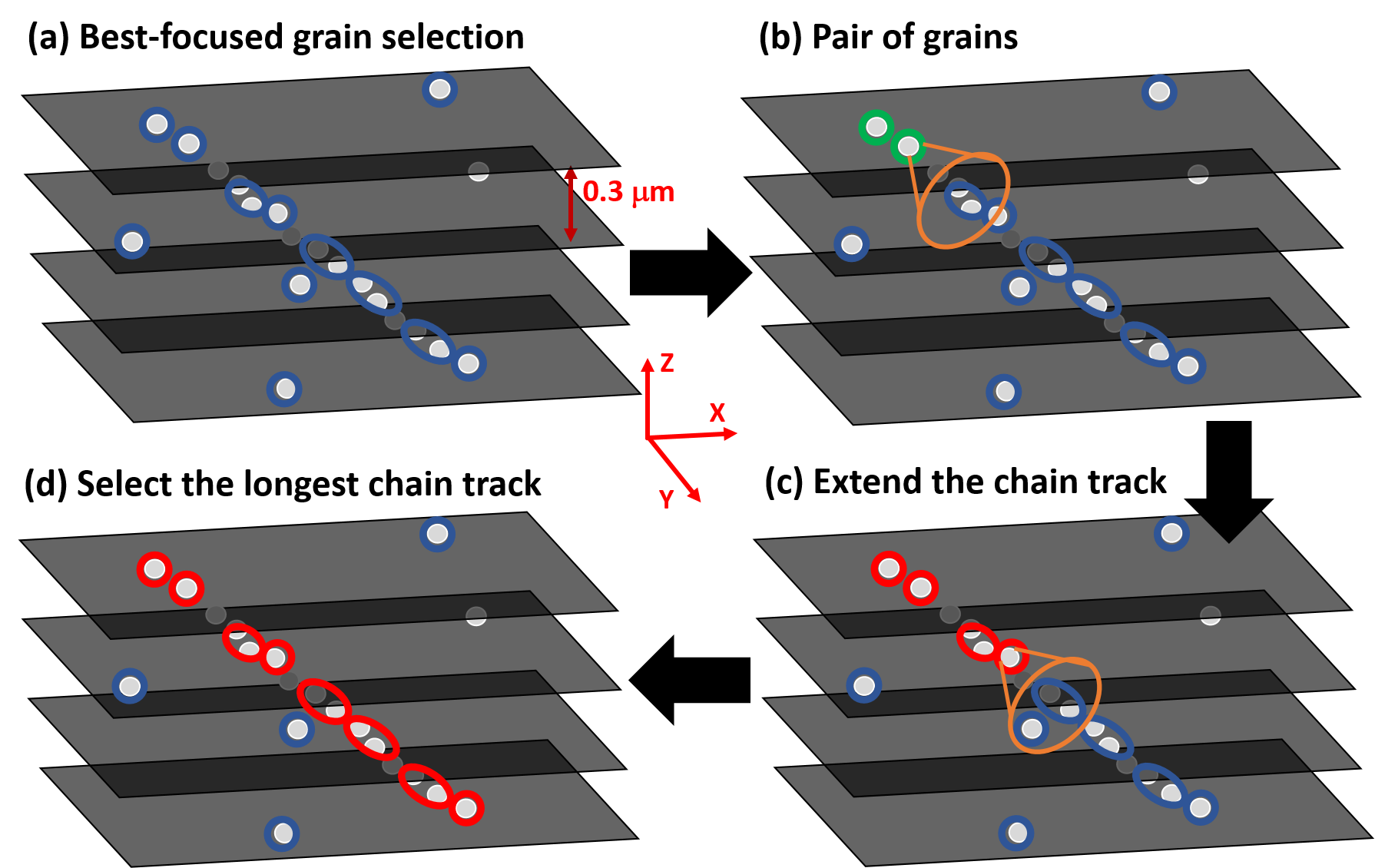}
  \caption{Algorithm of chain analysis. The blue clusters show the best-focused grains, the green ones show the pairs, and the red ones show the chain tracks.}
  \label{fig:algorithm}
\end{figure}

To analyze the high-resolution tomographic images acquired by PTS on SDRAM with high speed, we have developed a three-dimensional tracking system called chain analysis. This system can reconstruct three-dimensional tracks of more than a few micrometers length by connecting the developed silver grains like a chain, using the following procedures:

\begin{enumerate}
\item Image filtering such as smoothing and subtracting the background brightness are performed for each layer of the tomographic images, and the contour shapes of the developed silver, which are called grains, are extracted (Fig. \ref{fig:filtering}).
\item The unfocused grains are erased by selecting the best-focused grains from the top and bottom tomographic images (Fig. \ref{fig:algorithm}a).
\item Among the best-focused grains, those with a distance of less than 1.1 $\mu$m are connected to form a pair (Fig. \ref{fig:algorithm}b).
\item The angle of the pair is extended, and the pair is connected to another grain with a distance of less than 1.7 $\mu$m and an angle difference of less than 30 \degree (Fig. \ref{fig:algorithm}c).
\item All connections are attempted like a chain, and the longest one is selected as the representative track (Fig. \ref{fig:algorithm}d).
\item Among the detected chain tracks, those with low brightness or that are composed of grains larger than the developed silver are cut during offline analysis. In addition, a fiducial volume cut as in manual analysis is applied.
\end{enumerate}

\subsection{Performance of Chain Analysis}
\label{subsec:performance}

\begin{figure}[h!]
  \centering
  \includegraphics[width=16cm,bb=0 0 1100 600]{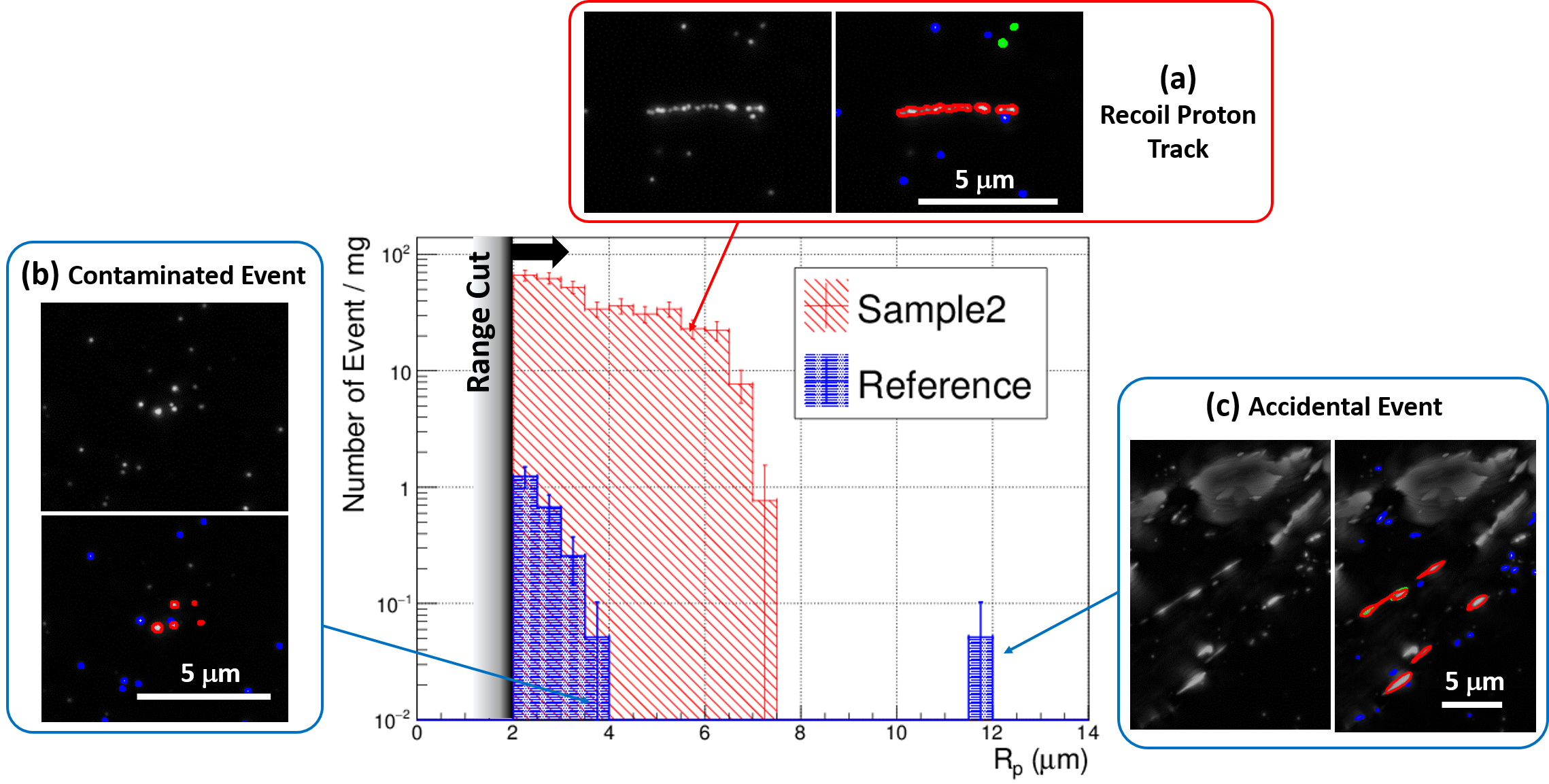}
  \caption{Track range distribution of candidate events obtained by chain analysis. Analyzed mass of 1.3 mg for Sample2, and 19.4 mg for reference sample were normalized to 1 mg. (a) is a recoil proton track triggered by chain analysis for Sample2. (b) and (c) are misrecognized noises for reference sample.}
  \label{fig:auto}
\end{figure}

The neutron-exposed samples were automatically analyzed using chain analysis, and its performance was evaluated. Fig. \ref{fig:auto} shows the range distribution of candidate events from chain analysis of a mass of 1.3 mg for Sample2. The same analysis of a mass of 19.4 mg for reference sample was performed to estimate the amount of noise misrecognized by chain analysis.

\begin{figure}[h!]
  \centering
  \includegraphics[width=12cm,bb=-100 0 700 400]{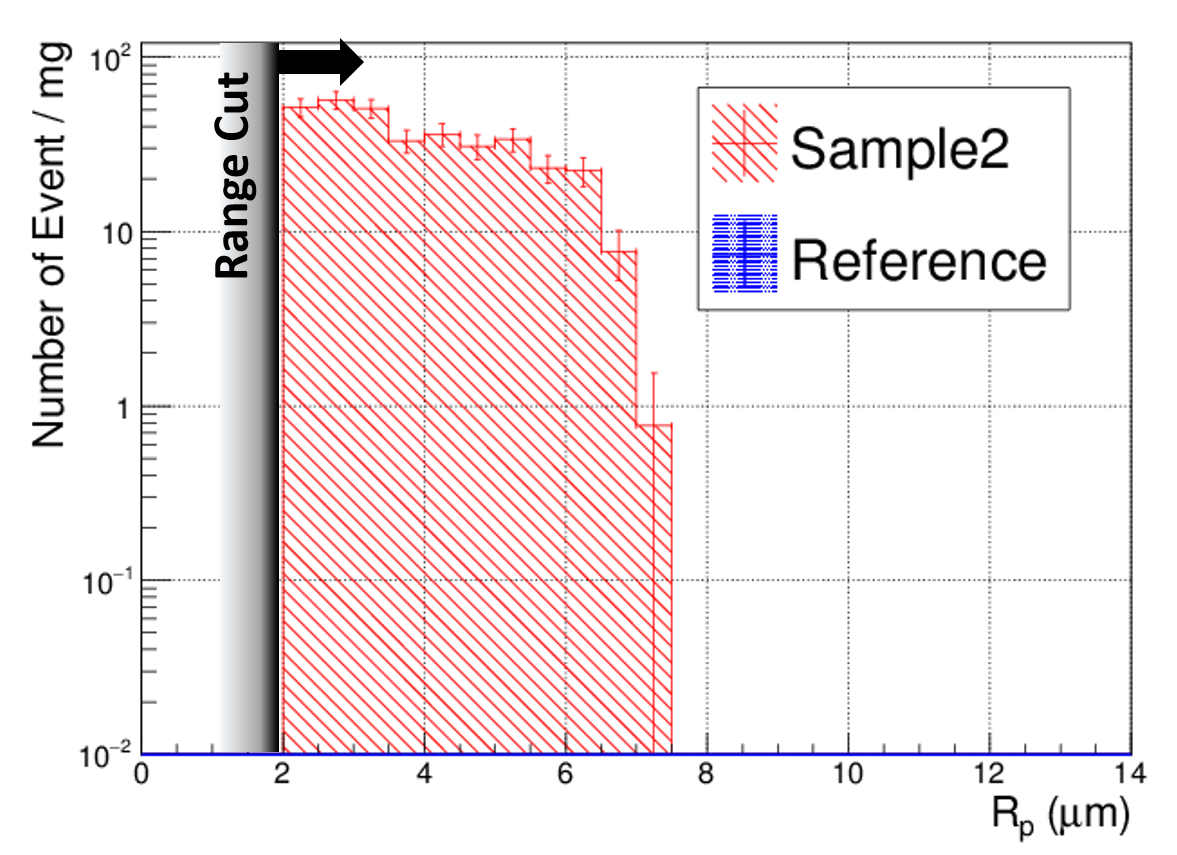}
  \caption{Track range distribution of proton tracks confirmed by manual analysis among events triggered by chain analysis. Analyzed masses of 1.3 mg for Sample2 and 19.4 mg for the reference sample were normalized to 1 mg.}
  \label{fig:auto_man}
\end{figure}

In neutron-exposed Sample2, a significant number of candidates triggered by chain analysis were detected against the reference sample, and most of them were identified as proton recoil tracks, as shown in the actual image (Fig. \ref{fig:auto}a). In addition, the same amount of noises as in the reference sample is present in the candidate events of Sample2. The main cause of noise is the connection of several small dust particles (Fig. \ref{fig:auto}b), and there are also a few other components misidentified the part of the huge dust particle (Fig. \ref{fig:auto}c). Such dust particles are predicted to be generated during the manufacturing process of the NIT. The noise density is 2.22 $\pm$ 0.34 event/mg in the case of the 2 $\mu$m cut, and 0.05 $\pm$ 0.05 event/mg in the case of the 4 $\mu$m cut. These noises are not an essential background, and they can be discriminated from tracks by manual analysis. Fig. \ref{fig:auto_man} shows the range distribution after manual analysis for the events triggered by chain analysis. Most of them were identified as proton tracks in Sample2; on the other hand, no tracks were observed in the reference sample. The number of events obtained for each analysis is summarized in Table \ref{tab:nevent}.

\begin{table}[htb]
  \caption{Number of candidates triggered by chain analysis, and number of proton tracks confirmed by manual analysis.}
  \centering
  \begin{tabular}{|c|c|c|c|} \hline
    Range cut & Sample & Candidates triggered by & After manual analysis \\
     & & chain analysis (/mg) & (/mg) \\ \hline \hline
    2 $\mu$m & Sample2 & 370 $\pm$ 17 & 348 $\pm$ 16 \\ \cline{2-4}
    (E$_{p} > $ 220 keV) & Reference & 2.32 $\pm$ 0.35 & $<$0.12 (90\% C.L.) \\ \hline
    4 $\mu$m & Sample2 & 155 $\pm$ 11 & 155 $\pm$ 11 \\ \cline{2-4}
    (E$_{p} > $ 380 keV) & Reference & 0.05 $\pm$ 0.05 & $<$0.12 (90\% C.L.) \\ \hline
  \end{tabular}
  \label{tab:nevent}
\end{table}

\begin{figure}[h!]
  \begin{minipage}[b]{0.5\linewidth}
    \centering
    \subfloat[]{\includegraphics[width=10cm,bb=0 0 800 450]{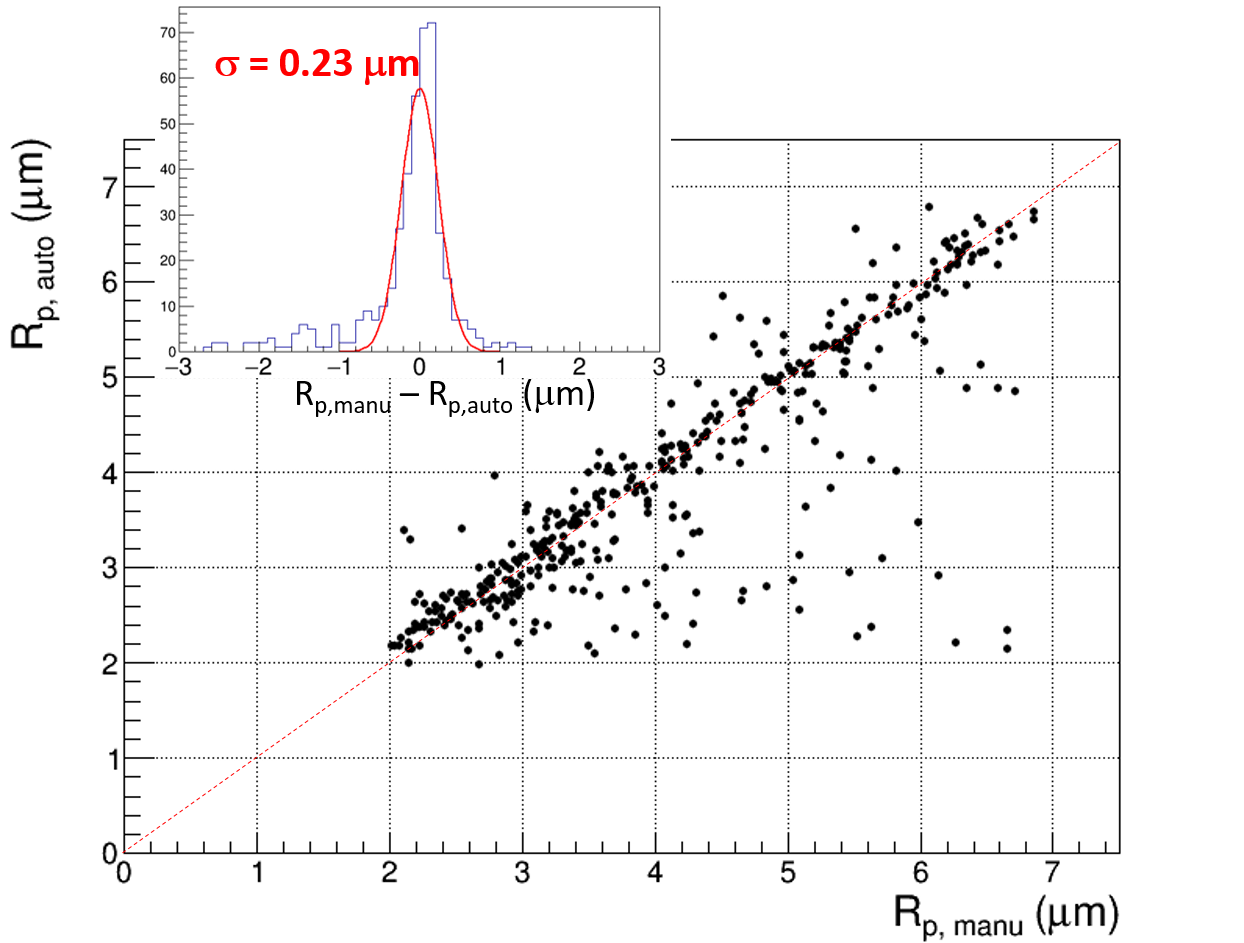}\label{fig:acc_range}}
  \end{minipage}
  \begin{minipage}[b]{0.5\linewidth}
    \centering
    \subfloat[]{\includegraphics[width=10cm,bb=0 0 800 450]{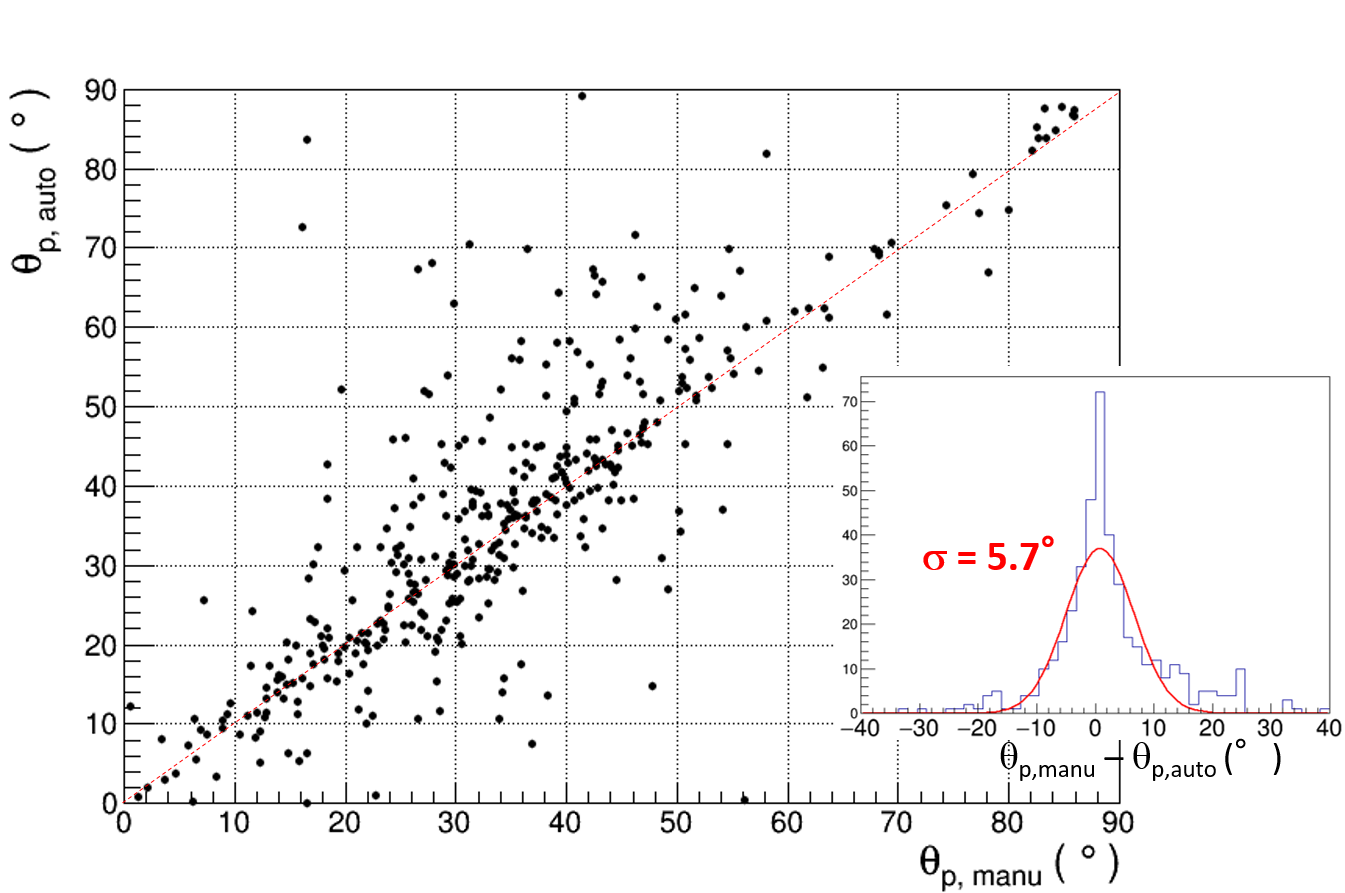}\label{fig:acc_theta}}
  \end{minipage}
  \begin{minipage}[b]{0.5\linewidth}
    \centering
    \subfloat[]{\includegraphics[width=10cm,bb=0 0 800 400]{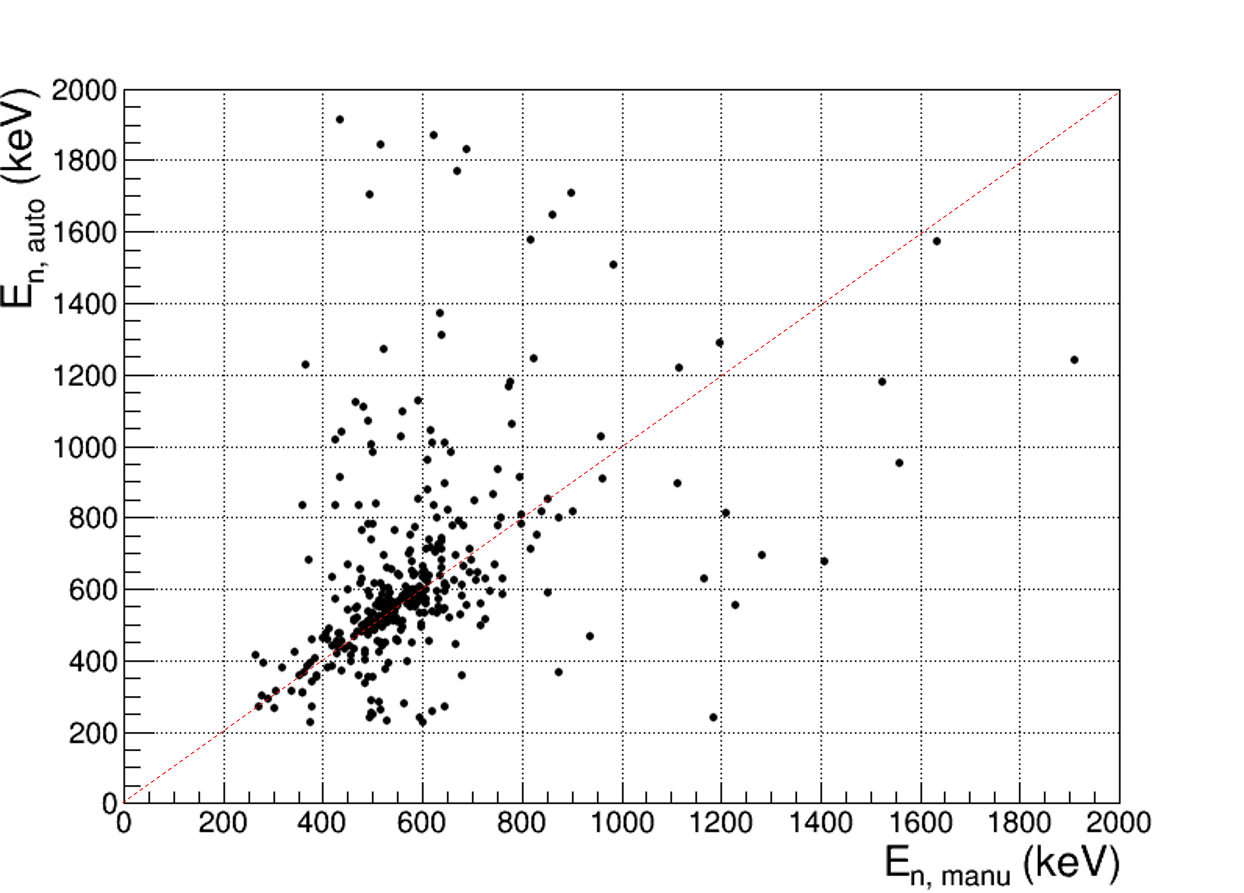}\label{fig:acc_En}}
  \end{minipage}
  \begin{minipage}[b]{0.5\linewidth}
    \centering
    \subfloat[]{\includegraphics[width=12cm,bb=0 0 800 400]{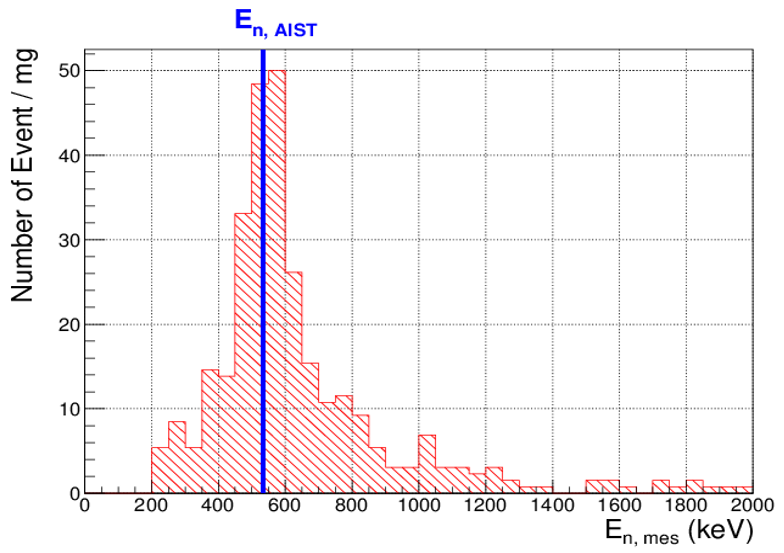}\label{fig:En_man_auto}}
  \end{minipage}  
  \caption{Performance evaluation of (a) track range, (b) scattering angle, and (c) reconstructed neutron energy by comparing manual analysis and chain analysis for Sample2.
  (d) Reconstructed neutron energy distribution with measurement accuracy of chain analysis.}
  \label{fig:auto_accuracy}
\end{figure}

The recoil protons detected by manual analysis and chain analysis are associated with each other, and the recognition accuracy for proton tracks in chain analysis is evaluated by a comparison with manual analysis.
As shown in Figs. \ref{fig:acc_range}, \ref{fig:acc_theta}, the recognition accuracy of the track range is 0.23 $\mu$m and that of the scattering angle is 5.7\degree. In chain analysis, the chain track may be broken when the distance between grains is large or when the brightness of the grain is low, and the track range may be recognized as shorter. Further, the recognition accuracy for the Z position is worse than that for the XY position due to the depth of field of the optical system, and therefore the angle accuracy in the Z direction is relatively poor. Figs. \ref{fig:acc_En}, \ref{fig:En_man_auto} show the result of the neutron energy reconstruction with the recognition accuracy of chain analysis. Sub-MeV neutron energies can be sufficiently reconstructed with the accuracy of automatic analysis.

\begin{figure}[h!]
  \centering
  \includegraphics[width=8cm,bb=0 0 800 600]{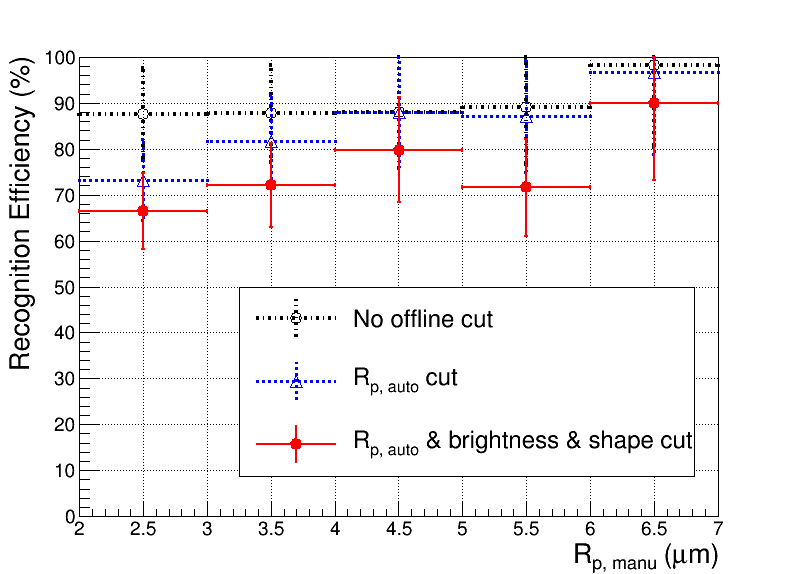}
  \caption{Recognition efficiency of proton tracks with chain analysis for Sample2. Horizontal axis corresponds to the proton track range measured by manual analysis. The black dotted line represents "without offline cut", the blue dotted line represents "after ${R}_{p,auto}$ cut", and the red line represents "after ${R}_{p,auto}$, brightness and shape cut".}
  \label{fig:auto_eff}
\end{figure}

Fig. \ref{fig:auto_eff} shows the recognition efficiency, which is defined as the relative efficiency of chain analysis to that of manual analysis, for proton tracks longer than 2 $\mu$m. The average recognition efficiency is approximately 90\%. However, by applying a cut for the proton range automatically measured by chain analysis with $>$2 $\mu$m (${R}_{p,auto}$ cut), it becomes (83 $\pm$ 4)\%. Furthermore, some other offline cuts such as brightness and shape are applied to distinguish noise, after which the recognition efficiency becomes (74 $\pm$ 4)\%.

\subsection{$\gamma$-ray Rejection Power}
\label{subsec:gamma}

%In sub-MeV environmental neutron detection, high rejection power for $\gamma$-rays is required because environmental $\gamma$-rays of similar energy can become the background.
In case of a high $\gamma$/neutron ratio environment, a high rejection power for $\gamma$-rays is required for sub-MeV neutron detection.
In the standard nuclear emulsions sensitive to minimum ionizing particles (MIPs), secondary electrons from the $\gamma$-rays are recorded as tracks. However, because the super-fine-grained AgBr:I crystals in the NIT drastically reduce the sensitivity to particles with small energy losses such as MIPs, most electrons cannot create LISs. Only the electrons immediately before stopping, with an energy of approximately 10 keV, can make a few LISs because their energy loss increases to approximately 10 keV/$\mu$m according to the Bethe-Bloch formula \cite{Bethe}.

To quantitatively evaluate the $\gamma$-ray rejection power of the NIT, we first considered the possibility that sub-MeV electrons recoiled by the Compton effect could be detected as tracks. Using $^{60}$Co as a $\gamma$-ray source with energies of 1.17 and 1.33 MeV, a NIT sample exposed to them with 10$^{7}$ $\gamma$/cm$^{2}$ flux, which is equivalent to one year's accumulation of environmental $\gamma$-rays, was prepared.
The expected number of Compton electrons in this sample is approximately 10$^{3}$ in the analyzed volume of 3.3 mg, as shown in Table \ref{tab:gamma_expected}. Chain analysis on this sample did not detect a significant electron signal, from the result, the detection efficiency of electron tracks is lower than 0.1\%, and secondary electrons generated from the $\gamma$-rays cannot be the background, considering the reaction probability.

Next, if LISs created by secondary electrons from low-energy $\gamma$-rays with a large cross section accumulate, chance coincidence events associated with them may be misrecognized by chain analysis. To evaluate this possibility, we also prepared a NIT sample exposed to 10$-$60 keV $\gamma$-rays from the $^{241}$Am source with 10$^{7}$ $\gamma$/cm$^{2}$ flux. The expected number of secondary electrons generated in this NIT sample is approximately 1$\times$10$^{5}$ in the analyzed mass volume of 3.3 mg. Chain analysis for this sample did not detect a significant signal, from the result, the number of chance coincidence events is less than the detection sensitivity even when secondary electrons generated from low-energy $\gamma$-rays accumulated.

This performance shows that $\gamma$-rays in any energy band cannot be the background when accumulated for less than one year.

\begin{table}[htb]
  \caption{Expected number of reactions in NIT samples exposed to $\gamma$-rays of $^{60}$Co and $^{241}$Am.}
  \centering
  \begin{tabular}{|c|c|c|c|} \hline
    $\gamma$-ray source & Energy & Exposed flux & Expected number of reacted $\gamma$-rays \\
    & (keV) & (/cm$^{2}$) & in analyzed volume of 3.3 mg \\ \hline
    $^{60}$Co  & 1170, 1330 & 10$^{7}$ & 10$^{3}$ \\ \hline
    $^{241}$Am & 10 $-$ 60 & 10$^{7}$ & 10$^{5}$ \\ \hline
  \end{tabular}
  \label{tab:gamma_expected}
\end{table}

\begin{comment}
\begin{figure}[h!]
  \begin{minipage}[b]{0.5\linewidth}
    \centering
    \subfloat[]{\includegraphics[width=11cm,bb=0 0 800 450]{gamma_Co_19.4mg.png}\label{fig:Co}}
  \end{minipage}
  \begin{minipage}[b]{0.5\linewidth}
    \centering
    \subfloat[]{\includegraphics[width=11cm,bb=0 0 800 450]{gamma_Am_19.4mg.png}\label{fig:Am}}
  \end{minipage}
  \caption{}
  \label{fig:gamma}
\end{figure}
\end{comment}

\section{Discussion and Prospects}

\subsection{Background}

As shown in section \ref{subsec:gamma}, because this neutron measurement technique has a high $\gamma$-ray rejection power, environmental $\gamma$-rays cannot be the background even after one year of accumulation. Dust events that are misrecognized by chain analysis also cannot be a fundamental background because it can easily be identified by manual analysis. However, because the capacity of manual analysis is limited, device purification is required when the analysis scale increases or when the current proton energy threshold (240 keV) decreases.

$\alpha$-rays should also be considered as the background when applying this analysis to low flux neutron measurement such as environmental neutron measurement. External $\alpha$-rays can be rejected with the current fiducial volume cut. We consider radioisotopes of $^{238}$U and $^{232}$Th included in the NIT as the intrinsic background; they are measured by germanium detectors as respectively 27 mBq/kg and 6 mBq/kg \cite{Activity}, and the number of $\alpha$-rays emitted by them is expected to be 8.6 $\times$ 10$^4$ event/(kg month). Because their track range exceeds 20 $\mu$m and that of recoil protons induced by 1 MeV neutrons is 15 $\mu$m, when we target sub-MeV neutron measurement, $\alpha$-rays can be rejected by the track range cut. Furthermore, because $\alpha$-rays have a larger dE/dx than protons, proton/$\alpha$ discrimination may be possible by the sensitivity control of AgBr:I crystals in the NIT.

\subsection{Environmental Neutron Measurement}

Because environmental neutrons, especially underground neutrons, can be background events in direct dark matter search and neutrinoless double beta-decay search experiments, a detailed understanding of environmental neutrons is important for making those discoveries. In particular, environmental neutrons in the sub-MeV region have not yet been directly measured due to technical difficulties.

A study has been conducted by A. Rindi et al. \cite{Neutron_GranSasso} on the measurement of environmental neutrons at the Gran Sasso Science Institute. The $^3$He proportional counter used in the study is well suited for the measurement of thermal neutrons. However, in the case of sub-MeV neutrons, because it detects protons by the neutron absorption reaction $^3$He(n,p) after deceleration by the moderator, it has no energy resolution, and the systematic error in the deceleration process is large. Furthermore, it has no directional sensitivity for neutrons because of the lack of spatial resolution.

\begin{comment}
\begin{table}[htb]
\centering
\caption{Analysis scale for sub-MeV environmental neutron measurement at Gran-Sasso.}
\begin{tabular}{|c|c|c|} \hline
  & Neutron Flux @ 0.1 - 1 MeV & Expected Proton Recoils \\
  & (cm$^{-2}$ s$^{-1}$) & in NIT (kg$^{-1}$ month$^{-1}$) \\ \hline \hline
  Surface & 3 $\times$ 10$^{-3}$ & 3 $\times$ 10$^5$ \\ \hline
  Underground & 10$^{-6}$ & 100 \\ \hline
\end{tabular}
\label{tab:env_neutron}
\end{table}
\end{comment}

Considering the neutron flux predicted by EXPACS \cite{EXPACS}, approximately 300 recoil proton tracks induced by environmental sub-MeV neutrons are accumulated by installing a 1 g NIT for 1 month on the Gran Sasso surface. Concerning the underground sub-MeV environmental neutron measurement, 1-kg-scale NIT analysis capability is required for a 1 month installation. The tracks recorded in the NIT will fade with time in high-temperature and high-humidity conditions. Because there is no significant fading within 1 month at 0\degreeC, we assume the current running period to be 1 month. With a longer running period, the long-term stability of the NIT should be confirmed at low temperature.

\subsection{Readout Speed}

The analysis speed for environmental neutron measurement with the NIT is limited by the image-taking speed of the PTS and image-processing speed of chain analysis. At present, an analysis speed of 30 g/year per PTS machine has been achieved. 1-g-scale analysis of the NIT can be performed in half a month with the one current PTS system.

For the underground neutron measurement, it is necessary to further improve the analysis speed. In the development stage, we have achieved an analysis speed of 85 g/(year $\cdot$ machine) using a wide F.O.V. objective lens with an optical magnification of 60$\times$. In addition, image filtering processes such as smoothing and subtracting background brightness are currently very time consuming, and the analysis speed can be improved to 150 g/(year $\cdot$ machine) by processing them with a GPU. Furthermore, to improve the image-taking speed of the PTS, we are planning to introduce a new high-speed wide-imaging CMOS camera, and a piezo device for high-speed driving. An analysis speed of scale of 1 kg/(year $\cdot$ machine) is expected with these improvements.

Currently, three PTS machines are in operation, and we plan to further increase the number of machines in the future.

%\subsection{Boosted Dark Matter}

\section{Conclusion}

We proposed a new sub-MeV neutron detection technique with NIT and PTS systems. We prepared a sample exposed to monochromatic neutrons at AIST, and evaluated the detection performance using manual analysis. The detection efficiency for recoil proton tracks was demonstrated to be 100\% consistent, and it was also shown that the energy of sub-MeV neutrons can be reconstructed from the recoil proton range and scattering angle.

We developed a new algorithm for automatically recognizing recoil proton tracks of more than a few micrometers for large-scale analysis. In this analysis, $\gamma$-rays are not detected as a background, which is an appropriate detection method with energy resolution for sub-MeV neutron measurement. In the future, we will continue to further improve the recognition method and speed up the analysis, and measure environmental neutrons both at the surface and underground.

\section*{Acknowledgment}
This work was supported by JSPS KAKENHI Grant Numbers JP18H03699, JP19H05806. Neutron Sources were supported by Dr. Tetsuro Matsumoto and Dr. Akihiko Masuda of the National Metrology Institute of Japan (NMIJ), the National Institute of Advance Industrial Science and Technology (AIST). The work is also supported  by JSPS Core- to-Core Program (grant number: JPJSCCA20200002).

\section*{Appendix}
First, the NIT film was treated with 0.49 mo/L sodium sulfate solution for 15 min as the pre-soak solution to stabilize the temperature of the film. This was followed by development treatment using MAA-1 Developer. The detailed prescription is shown in Table \ref{tab:MAA1}. This treatment was performed for 10 min. Stop and fixation treatment followed, and the solution utilized was acetic acid and NF-1 of Fujifilm. These treatments were performed for 10 min and 20 min, respectively. All processes were conducted at 5.0 \degreeC.
\begin{table}[htb]
\centering
\caption{Prescription of MAA-1 Developer}
\begin{tabular}{|c|c|} \hline
  p-methylaminophenol hydrochloride & 2.5 g \\
  L(+)-ascorbic acid  & 10 g \\ 
  Potassium bromide & 1 g \\
  Sodium metaborate tetrahydrate & 47.2 g \\
  Water to make &  1000 mL \\\hline
\end{tabular}
\label{tab:MAA1}
\end{table}

%\bibliography{references}
%\bibliographystyle{unsrt}

\end{document}